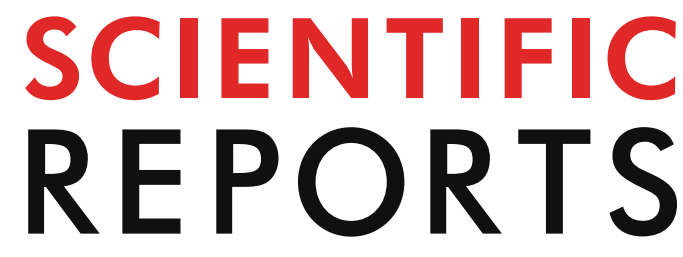

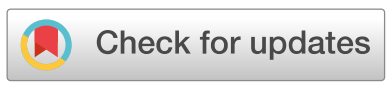

OPEN

# Potential quality improvement of stochastic optical localization nanoscopy images obtained by frame by frame localization algorithms

Yi Sun

**A data movie of stochastic optical localization nanoscopy contains spatial and temporal correlations, both providing information of emitter locations. The majority of localization algorithms in the literature estimate emitter locations by frame-by-frame localization (FFL), which exploit only the spatial correlation and leave the temporal correlation into the FFL nanoscopy images. The temporal correlation contained in the FFL images, if exploited, can improve the localization accuracy and the image quality. In this paper, we analyze the properties of the FFL images in terms of root mean square minimum distance (RMSMD) and root mean square error (RMSE). It is shown that RMSMD and RMSE can be potentially reduced by a maximum fold equal to the square root of the average number of activations per emitter. Analyzed and revealed are also several statistical properties of RMSMD and RMSE and their relationship with respect to a large number of data frames, bias and variance of localization errors, small localization errors, sample drift, and the worst FFL image. Numerical examples are taken and the results confirm the prediction of analysis. The ideas about how to develop an algorithm to exploit the temporal correlation of FFL images are also briefly discussed. The results suggest development of two kinds of localization algorithms: the algorithms that can exploit the temporal correlation of FFL images and the unbiased localization algorithms.**

In stochastic optical localization nanoscopy—PALM[1], STORM[2], FPALM[3] and (d)STORM[4], a localization nanoscopy image is produced by three steps. First, a set of emitters are attached to ultrastructure of a specimen. Second, in each frame time a random subset of emitters are activated by a laser and emit photons that pass through an optical lens and produce a data frame acquired by a camera. Repeating this process, a data movie that consists of a large number of data frames is acquired. Third, a localization algorithm estimates the emitter locations from the data movie and produces a localization nanoscopy image of the specimen ultrastructure. The localization algorithm plays an important role in obtaining a high quality of localization nanoscopy images.

In the past decade, a number of localization algorithms have been developed in the literature on the basis of a variety of criteria and objectives, including but not limited to *localization of single emitters in single frames*: (d)STORM[4], Octan[5], FluoroBancroft[6], Gaussian fitting[7], PeakSelector[8], SOFI[9], DAOSTORM[10], maximum likelihood[11], and palm3d[12], *localization of multiple emitters in single frames*: 3D-DAOSTORM[13], compressed sensing[14], fast maximum likelihood[15], RadialSymmetry[16], PeakFit[17], PALMER[18], RapidSTORM[19], least-square fitting with the 3D Gibson-Lanni point spread function (PSF)[20], PC-PALM[21], fast compressed sensing[22], Easy-DHPSF[23], 3D-WTM[24], RainSTORM[25], WaveTracer[26], µManager[27], ThunderSTORM[28], FALCON[29], MIATool[30], AO-STORM[31], state space[32], TVSTORM[33], ADCG[34], Cspline[35], ALM[36], SMAP[37], UNLOC[38], sparse Bayesian learning[39], LSTR[40], FCEG[41], WinSTORM[42], and QC-STORM[43], *localization of multiple emitters in multiple frames*: 3B analysis[44], deconSTORM[45], spatiotemporal decomposition and association[46], and nonnegative matrix factorization[47]. In addition, the recent approaches include cloud computing[48], clustering analysis[49,50], big

Electrical Engineering Department, Nanoscopy Laboratory, The City College of City University of New York, New York, NY 10031, USA. email: ysun@ccny.cuny.edu

data analysis[51], correlation analysis[52], HAWK[53] to alleviate artifacts detectable by the Fourier ring correlation (FRC)[54,55], neural networks[56,57], machine learning[58], and deep learning[59–61].

To boost research and development of localization algorithms and identify high-performance localization algorithms, an online public challenge has been open to the public[62–64]. The results of challenge on the 2D and 3D imaging have been reported in Ref.[63] and recently in Ref.[64]. In the challenge, a data movie is synthesized with a set of emitters whose locations are known. A localization algorithm estimates the emitter locations by using the data movie and produces a localization nanoscopy image consisting of the estimated emitter locations. The quality of the localization nanoscopy image and the performance of the algorithm are evaluated by comparison of the estimated emitter locations and the true emitter locations in terms of the quality metrics of accuracy, precision, recall, and Jaccard index[26,47,62,63]. A universal and objective metric, root mean square minimum distance (RMSMD), is proposed recently[66]. RMSMD measures the average, local, and mutual fitness between a set of estimated emitter locations and the set of true locations, and it presents several unique properties and advantages over the other metrics. RMSMD can be utilized to evaluate the quality of nanoscopy localization images and the performance of localization algorithms when a set of true emitter locations is known a prior[66].

Among the localization algorithms in the literature, only a few exploit temporal correlation by jointly utilizing multiple data frames or the entire data movie in estimation of emitter locations[44–47]. The majority of localization algorithms[1–43] estimate emitter locations from each single data frame independently or by the frame-by-frame localization (FFL). Thus, most localization nanoscopy images are FFL images. Yet, little is known about their properties. It is imperative and important in both theory and application to understand the properties of FFL nanoscopy images in several aspects. First, since the optical lens is effectively a PSF, a data frame is spatially (pixel-wise) correlated. Moreover, because all data frames are generated by the same set of emitters, the data movie is also temporarily (frame-wise) correlated. The Fisher information matrix of multiple data frames is equal to the sum of their individual Fisher information matrices[69]. Both the spatial and temporal correlations contain information about the emitter locations. If the spatial and temporal correlations are jointly and optimally exploited in localization of emitters, the localization accuracy can approach the bound that the data movie can provide. However, such an advanced localization algorithm is usually computationally complicated; and this is probably the reason why the majority of localization algorithms estimate emitter locations frame by frame independently. The FFL algorithms only exploit the spatial correlation and leave the temporal correlation to be intact. The temporal correlation is still contained in the FFL image, which if exploited, shall improve the localization accuracy of estimated emitter locations and the quality of nanoscopy image as well. The algorithms UNLOC[38] and QC-STORM[43] detect and reconnect the estimated locations that might be generated from the same emitter in consecutive frames in order to improve localization accuracy. It is interesting to know the potentially maximum improvement of quality that can be obtained by exploitation of the temporal correlation in an FFL image. Second, as the number of data frames increases, the number of activations per emitter in the data movie increases and then the number of estimated locations per emitter in an FFL image increases. It is interesting to know how the average number of estimated locations per emitter affects the quality of an FFL image. It is practically interesting to know if it is necessary to acquire as many data frames as possible in order to improve the quality of an FFL image and when an acquisition of data frames shall terminate. A structure-resolving index (SRI) has been proposed to determine the termination time of acquisition[70]. Third, it is desired to know how the variance and bias of localization errors and sample drift affect the quality of an FFL image. The effect of biases of localization errors has been recently paid attention in literature[65]. Understanding the effect of variance and bias of localization errors and sample drift on the image quality enables algorithm developers to allocate resources more adequately to achieve a high quality of FFL images. Fourth, serval deterministic properties of RMSMD are analyzed and presented in Ref.[66]. However, its statistical properties are unknown yet while a data movie and an FFL image are random realizations of certain stochastic processes. An analysis of RMSMD for FFL nanoscopy images shall reveal statistical insights and understandings of RMSMD[71]. Fifth, root mean square error (RMSE) can be utilized to theoretically analyze and quantify the error between the estimated localizations and the ground-truth locations of emitters. The Cramer-Rao lower bound (CRLB)[67,68] determines the minimum RMSE for all unbiased estimators. While RMSMD can be employed to evaluate the quality of a localization nanoscopy image in comparison with a set of ground-truth locations, RMSE cannot be employed in general since it needs to know the identification of an estimated location associated with a ground-truth location, which is unknown in general. Hence, it is significant to reveal the relationship between RMSMD and RMSE. Moreover, it is also interesting to know the properties of FFL nanoscopy images in terms of RMSE.

In this paper the statistical properties of RMSMD and RMSE for the FFL nanoscopy images are analyzed and compared. First, it is found that while an FFL image is random, its RMSMD converges to a deterministic constant as the average number of activations per emitter $\lambda$ tends to infinity. This implies that for a sufficiently large $\lambda$, increasing the number of acquired data frames improves little the quality of an FFL image in terms of reduction of RMSMD variation. A numerical example shows that when $\lambda = 10$, RMSMD is already stable and close to the limit RMSMD and acquiring more data frames is unnecessary. Second, the analytical result shows that exploitation of temporal correlation in an FFL image can reduce RMSMD and RMSE by a maximum fold of $\lambda^{0.5}$. Hence, an algorithm that is able to exploit the temporal correlation in an FFL image can significantly improve the image quality, in particular for a large $\lambda$. A numerical example shows that exploitation of temporal correlation not only reduces RMSMD and RMSE of an FFL image but also considerably improves its visual quality. Third, the variance of localization biases across emitters affects RMSMD and RMSE much more severely than the variance of localization errors. On the basis of the first two results, we can conclude that if only an FFL algorithm is available, acquiring more data frames is unnecessary when the average number of activations per emitter already reaches $\lambda = 10$. On the other hand, if an FFL algorithm is followed by an algorithm that can exploit the temporal correlation, acquiring more data frames can significantly improve the image quality in both RMSMD and visual quality. At the end, the ideas about how to develop an algorithm to exploit the temporal correlation of FFL images are also briefly discussed. These results suggest that in order to achieve a high quality of localization nanoscopy images it is important to develop two kinds of algorithms: the algorithms that can exploit temporal correlation contained in FFL images and the unbiased localization algorithms.

## Method

**FFL image, RMSMD and RMSE.** *FFL image.* Let $S = \{s_1, ..., s_M\}$ be a set of $M$ fixed emitter locations in the $n$-dimensional real space $\mathbb{R}^n$. In practice, the dimension is $n = 2$ for the 2D imaging and $n = 3$ for the 3D imaging. In a data movie of $L$ frames each emitter is independently activated, following a Markov chain as considered in the literature[44,45,66]. The Markov chain is irreducible, aperiodic, and positive recurrent and therefore has a stationary probability distribution of states[66] such that an emitter in a frame is activated with a stationary probability $p$. In practice, an emitter will be ultimately bleached. Without loss of generality and effect on results, we consider that the emitters are not bleached yet in the data movie. The $i$th emitter is activated $N_i < L$ times in the data movie. An FFL algorithm localizes by estimation the activated emitters in each single frame independently. Let $X_i = \{x_{i1}, ..., x_{iN_i}\}$ be the set of all the locations estimated in different frames for the $i$th emitter $s_i$. $X = \bigcup_{i=1}^{M} X_i$ consists of $N = \sum_{i=1}^{M} N_i$ locations for all $M$ emitters estimated from the data movie.

As an estimate of $S$, $X$ is an FFL nanoscopy image for $S$. In this paper, we do not investigate how to obtain $X_i$'s; instead, given $X_i$'s, we analyze the quality of $X$ and the potential quality improvement by exploitation of temporal correlation that is embedded in $X$. The quality of $X$ can be measured by the RMSMD and the RMSE between $X$ and $S$. We shall analyze the statistical properties of the RMSMD and the RMSE when the number of data frames $L$ is large.

*Statistics.* The $i$th emitter $s_i$ is activated in $N_i$ frames. The estimated locations of $s_i$ from the $N_i$ frames, $x_{ij} \in X_i$ for $j = 1, \ldots, N_i$, can be written as $x_{ij} = s_i + b_i + w_{ij}$. $b_i$ is the localization bias that is fixed in all estimates and $b_{ik}$ for $k = 1, \ldots n$, the $k$th component of $b_i$, is the bias of $x_{ijk}$, the $k$th component of $x_{ij}$. $w_{ij}$ is the localization error with zero mean $E(w_{ij}) = 0$ and varies randomly in different frames. The localization errors are resulted from the photon emissions of emitters, background autofluorescence, Gaussian noise, and an FFL algorithm, which all are mutually independent[67–69]. In stochastic optical localization nanoscopy, the photon emissions from emitters and background autofluorescence both are Poisson processes. It follows from the property of classifying a Poisson number of events[72] that the numbers of detected photons in pixels of different frames are independent. The thermal noise in electronic circuit is additive white Gaussian noise (AWGN)[68,73] and therefore the Gaussian noises in pixels of different frames are independent. Thus, given the activations of emitters, the data frames are mutually independent[69]. An FFL algorithm estimates the locations of activated emitters frame by frame independently. Hence, the localization errors of the same emitter in different frames are independent.

The localization errors of different emitters in the same frame are dependent if their PSFs are overlapped. Consequently, the covariance matrix of $w_{ij}$ varies with $j$ depending on the combination of activated emitters associated with $j$. For a total of $M$ emitters, the number of possible combinations of activated emitters in a frame is $2^M$. If considering only the emitters whose PSFs can be possibly overlapped locally, $M$ is small and the total number of combinations might be small in an experiment. In the condition that the $i$th emitter is activated, the total number of combinations of activated emitters in a frame, denoted by $c_{il}$, is equal to $J = 2^{M-1}$. Since the Markov chain of activations of an emitter is stationary and has a stationary probability distribution of state, the $l$ th combination $c_{il}$ for $l = 1, \cdots, J$ occurs with a stationary probability $q_{il}$. Denote by $\sigma_{ilk}^2 = E(w_{ijk}^2|c_{il})$ the variance of the $k$th component of $w_{ij}$ in the condition that the $l$th combination $c_{il}$ occurs. It is clear that $\sigma_{ilk}^2$ is stationary, that is, $\sigma_{ilk}^2$ depends on the occurrence of the $l$ th combination $c_{il}$ and is independent of the frame index distinguished by $j$. Therefore, the localization error of the $i$th emitter $w_{ij} \in X_i$ for $j = 1, \ldots, N_i$ is stationary, and thus the variance of its $k$th component is equal to $\sigma_{ik}^2 = E(w_{ijk}^2) = \sum_{l=1}^{J} q_{il} E(w_{ijk}^2|c_{il}) = \sum_{l=1}^{J} q_{il}\sigma_{ilk}^2$. Then the estimated locations $x_{ij}$ of the $i$th emitter have a stationary probability density function $f_i(x)$.

The total number of activations of the $i$ th emitter, $|X_i| = N_i$, has a mean $\lambda = pL$. $N$, the total number of estimated locations for all emitters in $X$, has a mean $M\lambda = MpL$. In practice, the total number of frames $L$ is statistically large. To theoretically analyze the property of the RMSMD for a large $L$, so called a large data behavior, we consider that $L$ tends to infinity. Therefore, the mean of all $N_i$'s as well as $N$ tend to infinity, i.e., $\lambda \to \infty$.

*RMSMD.* Given $S$ and $X$, their mean square minimum distance (MSMD) is defined by[66]

$$D^2(X, S) = \frac{1}{|X| + |S|}\left(\sum_{s\in S}\min_{x\in X}\|x - s\|^2 + \sum_{x\in X}\min_{s\in S}\|s - x\|^2\right) \tag{1}$$

where $|\cdot|$ is the number of elements in a set and $\|\cdot\|$ is the $l_2$ norm or the Euclidean distance between two points. Then the RMSMD is $D(X, S)$. As a universal and objective metric, $D(X, S)$ evaluates how well the two sets $X$ and $S$ averagely, locally, and mutually fit to each other.

In localization nanoscopy $X$ is random and so is $D(X, S)$. In other words, an FFL image obtained in practice is one realization of $X$. $D(X, S)$ can be applied to a particular realization of $X$.

The Voronoi cell of $s_i \in S$ is defined by $V(s_i) = \{x \in R^n, \|x - s_i\| \le \|x - s_j\|, j \ne i\}$. The Voronoi cell $V(x_i)$ for $x_i \in X$ is defined similarly. In terms of the Voronoi cells, the RMSMD can be expressed as

$$D^2(X, S) = \frac{1}{|X| + |S|}\left(\sum_{x\in X}\sum_{s\in S\bigcap V(x)}\|x - s\|^2 + \sum_{s\in S}\sum_{x\in X\bigcap V(s)}\|s - x\|^2\right). \tag{2}$$

*RMSE.* The mean square error (MSE) between $X$ and $S$ is defined as

$$h^2(X,S) = \frac{1}{M}\sum_{i=1}^{M}\frac{1}{N_i}\sum_{j=1}^{N_i}E\left(\|x_{ij}-s_i\|^2\right) \tag{3}$$

where the expectation is taken with respect to $f_i(x)$'s. Then their RMSE is given by $h(X,S)$. It is shown in "Appendix" that the MSE can be expressed in terms of the variances and biases of all estimated emitter locations

$$h^2(X,S) = \frac{1}{M}\sum_{i=1}^{M}\sum_{k=1}^{n}\left(\sigma_{ik}^2 + b_{ik}^2\right). \tag{4}$$

In general, RMSMD and RMSE are quite different. First, RMSMD is random while RMSE is deterministic. Second, given $X$ and $S$, $D(X,S)$ can be employed to evaluate the quality of a localization nanoscopy image $X$. RMSMD is useful in practice as well as theoretical analysis. Third, in contrast, evaluation of RMSE need to know the partition $X_i$'s of the image $X$ and the probability density function of estimated locations $f_i(x)$. Therefore, RMSE is useful only in theoretical analysis. It is noted that the CRLB[67,68] is the minimum standard deviation $\sigma_{ijk}$ that any unbiased (i.e., $b_{ik}=0$ for all $k$) estimator $x_{ijk}$ can possibly achieve. Though they are quite different, as analyzed in the next section, in the special cases when $X$ is sufficiently dense and all the estimated locations are located inside the Voronoi cells of their own emitters, the random RMSMD can well approximate the deterministic RMSE.

## Properties.

*Invariance to a large number of estimates.* Since an FFL image is random, its RMSMD is random. However, as the average number of activations per emitter increases, the randomness of RMSMD is eventually effectively ignorable as indicated by the following property, which is proved in "Appendix".

*Property 1 (Invariance)*: As $\lambda \to \infty$, the RMSMD between $X$ and $S$ converges almost surely as

$$\lim_{\lambda\to\infty} D^2(X,S) = E\left(\min_{s\in S}\|s-x\|^2\right) \tag{5}$$

$$= \frac{1}{M}\sum_{i=1}^{M}\sum_{j=1}^{M}\int_{V(s_i)}\|s_i-x\|^2 f_j(x)dx \tag{6}$$

where the expectation in Eq. (5) is taken with respect to $f_i(x)$'s. □

In practice, for a large and finite $\lambda$, Eqs. (5) and (6) provide an approximation of RMSMD for an FFL image. In particular, Eq. (6) implies that

$$D^2(X,S) \cong \frac{1}{N}\sum_{i=1}^{M}\sum_{x\in X\bigcap V(s_i)}\|s_i-x\|^2 \tag{7}$$

In a particular experiment, an FFL image is a realization of random $X$ and might have a much poorer quality than the average in terms of RMSMD. As indicated by Property 1, however, as $\lambda \to \infty$, the random $D^2(X,S)$ converges to a deterministic constant at the right-hand side of Eq. (5). For a sufficiently large $\lambda$, the quality of $X$ in terms of $D(X,S)$ in any experiment shall be almost the same. This implies that if $\lambda$ is sufficiently large, further increasing $\lambda$ does not decrease the variation of RMSMD and acquiring more data frames is unnecessary. The numerical example in the next section shows that when the average number of activations per emitter reaches $\lambda = 10$, the RMSMD is already steady with small variations.

The RMSMD formulas in Property 1 are applicable to all cases of localization errors for FFL images. In the following two subsections, the RMSMD is further analyzed in the cases of small and large localization errors.

*RMSMD in small localization errors.* In general, the random RMSMD and the deterministic RMSE are irrelative. However, in the special case when $\lambda$ is sufficiently large and localization errors are small, the RMSMD effectively coincides with the RMSE. We consider that all locations $x_{ij}$'s estimated by an FFL algorithm are located in the Voronoi cell of their own emitter location $s_i$ with probability one, that is, $\Pr\left(X_i \bigcap V(s_j) = X_i\right) = \delta_{ij}$ for all $i$, $j$ where $\delta_{ij}$ is the Kronecker delta. This implies that the localization errors measured by RMSE are relatively small compared with the Voronoi cells. The following property is proved in "Appendix".

*Property 2 (Coincidence with RMSE)*: If $x_{ij}$'s all are located in the Voronoi cells of their own emitter locations with probability one, then in the almost sure sense

$$\lim_{\lambda\to\infty} D^2(X,S) = h^2(X,S), \tag{8}$$

which is equal to the right-hand side of Eq. (4). □

In an experiment, if most estimated locations $x_{ij}$'s are in the Voronoi cells of their own $s_i$'s and $\lambda$ is sufficiently large, then $D^2(X,S) \cong h^2(X,S) = M^{-1}\sum_{i=1}^{M}\sum_{k=1}^{n}(\sigma_{ik}^2 + b_{ik}^2)$.

***RMSMD upper bound in large localization errors.*** Consider that an FFL image is defined over a finite region $\Omega \subset \mathbb{R}^n$ and the variances of all estimated locations $X_{ij}$'s and the average number of activations per emitter both tend to infinity, i.e., $\sigma_{ik}^2 \to \infty$ and $\lambda \to \infty$. Then infinitely many estimated locations are uniformly distributed in $\Omega$. The uniform distribution of estimated locations is equivalent to a random guess of the emitter locations and no information about the emitter locations is contained in such a localization nanoscopy image. Because of this, the RMSMD of this limit image can be considered the upper bound of RMSMD, which correspond to the worst quality of a localization nanoscopy image. The following property is straightforward to obtain from (6).

*Property 3 (Upper bound on large localization error)*: As $\sigma_{ik}^2 \to \infty$ for all $i, k$ and $\lambda \to \infty$, in the almost sure sense

$$\lim_{\{\sigma_{ik}^2 \to \infty\}} \lim_{\lambda \to \infty} D^2(X, S) = \frac{1}{M} \sum_{i=1}^{M} \frac{1}{|V(s_i) \cap \Omega|} \int_{V(s_i) \cap \Omega} \|s_i - x\|^2 dx \tag{9}$$

where $|\cdot|$ denotes the volume of a continuous set. □

*Exploitation of temporal correlation.* We consider exploitation of temporal correlation retained in an FFL image and investigate the RMSMD improvement that is achievable by exploitation of temporal correlation.

The locations $x_{ij} \in X_i$ are estimates for the same emitter location $s_i$ from different frames. When the distribution of the localization errors is unknown, the statistical property of the errors cannot be exploited. However, when the localization errors of the same emitter across frames are independent, an efficient and simple method to improve the accuracy of estimated emitter location is the averaging of the estimates $x_{ij} \in X_i$ for $s_i$ to cancel out part of the errors, that is

$$\hat{x}_i = \frac{1}{N_i} \sum_{j=1}^{N_i} x_{ij}. \tag{10}$$

For the mutually independent localization errors of $x_{ij}$'s, $\hat{x}_i$ is the linear minimum-mean-square-error (LMMSE) estimator of the emitter location and bias. If the localization errors are further Gaussian distributed, $\hat{x}_i$ is the maximum likelihood estimator[74]. In these senses, when the distribution of localization errors is unknown, $\hat{x}_i$ is an optimal estimator to exploit the temporal correlation from an FFL image.

Now, let $\hat{X}_i = \{\hat{x}_i\}$ and $\hat{X} = \bigcup_{i=1}^{M} \hat{X}_i$; and then $\hat{X}$, as an estimate of $S$ that has the same number $M$ of emitter locations as that of $S$, is a new localization nanoscopy image. The following formula is shown in "Appendix",

$$\lim_{\lambda \to \infty} \lambda \left( h^2(\hat{X}, S) - \frac{1}{M} \sum_{i=1}^{M} \sum_{k=1}^{n} b_{ik}^2 \right) = \frac{1}{M} \sum_{i=1}^{M} \sum_{k=1}^{n} \sigma_{ik}^2. \tag{11}$$

The averaging of estimated locations estimated for the same emitter reduces the MSE on the part of variance by a fold of $\lambda$ but does not do on the part of biases. In the limit, the effect of error variances vanishes and the MSE

$$\lim_{\lambda \to \infty} h^2(\hat{X}, S) = \frac{1}{M} \sum_{i=1}^{M} \sum_{k=1}^{n} b_{ik}^2 \tag{12}$$

is determined only by the biases.

The following property is proved in "Appendix".

*Property 4 (Averaging)*: If $x_{ij}$'s all are located in the Voronoi cells of their own emitter locations with probability one, then in the almost sure sense

$$\lim_{\lambda \to \infty} D^2(\hat{X}, S) = \lim_{\lambda \to \infty} h^2(\hat{X}, S), \tag{13}$$

which is equal to the right-hand side of Eq. (12). □

Property 1–Property 4 imply that in the large data limit, exploitation of temporal correlation by the averaging of estimated locations per emitter can improve RMSMD by a fold of

$$\lim_{\lambda \to \infty} \frac{D^2(X, S)}{D^2(\hat{X}, S)} = \sum_{i=1}^{M} \sum_{k=1}^{n} (\sigma_{ik}^2 + b_{ik}^2) / \sum_{i=1}^{M} \sum_{k=1}^{n} b_{ik}^2. \tag{14}$$

In an experiment, if most estimated locations $x_{ij}$'s are in the Voronoi cells of their own $s_i$'s and $\lambda$ is sufficiently large, $D^2(\hat{X}, S)$ is approximately by Eqs. (11) and (13)

$$D^2(\hat{X}, S) \cong \frac{1}{M} \sum_{i=1}^{M} \sum_{k=1}^{n} \left( \frac{\sigma_{ik}^2}{\lambda} + b_{ik}^2 \right). \tag{15}$$

Hence, the RMSMD is improved by an approximate fold of

$$\frac{D^2(X, S)}{D^2(\hat{X}, S)} \cong \sum_{i=1}^{M} \sum_{k=1}^{n} (\sigma_{ik}^2 + b_{ik}^2) / \sum_{i=1}^{M} \sum_{k=1}^{n} \left( \frac{\sigma_{ik}^2}{\lambda} + b_{ik}^2 \right), \tag{16}$$

which converges to the right-hand side of Eq. (14), the maximum fold of improvement for biased estimates.

*Maximum fold of improvement in RMSMD.* The improvement of RMSMD in Eq. (16) is limited by the bias. If $x_{ij}$'s are all unbiased with $b_i = 0$, the improvement tends to infinity as $\lambda$ increases. We investigate the rate of RMSMD improvement.

With the unbiased estimates $x_{ij}$'s, Eqs. (4) and (11) become, respectively

$$h^2(X,S) = \frac{1}{M}\sum_{i=1}^{M}\sum_{k=1}^{n}\sigma_{ik}^2, \tag{17}$$

$$\lim_{\lambda\to\infty} \lambda h^2(\hat{X},S) = h^2(X,S). \tag{18}$$

The following property is obtained by means of Property 2 and Property 4.

*Property 5 (Maximum fold of improvement)*: If $x_{ij}$'s all are unbiased and with probability one are located in the Voronoi cells of their own emitter locations, respectively,

$$\lim_{\lambda\to\infty} \lambda D^2(\hat{X},S) = \lim_{\lambda\to\infty} D^2(X,S), \tag{19}$$

which is equal to the right-hand side of Eq. (17). □

Property 5 implies that as $\lambda \to \infty$, $D(\hat{X},S) \to 0$ at the rate of $\lambda^{-0.5}$. In practice, RMSMD shall be in the order of

$$D^2(\hat{X},S) \cong \frac{1}{\lambda M}\sum_{i=1}^{M}\sum_{k=1}^{n}\sigma_{ik}^2. \tag{20}$$

The averaging of estimated locations per emitter can improve RMSMD by a fold of

$$\frac{D(X,S)}{D(\hat{X},S)} \cong \sqrt{\lambda}. \tag{21}$$

In practice, $x_{ij}$'s are usually not located in their own Voronoi cells and all estimated locations are mingled together. Moreover, $x_{ij}$'s are usually biased. Furthermore, an algorithm that determines the partition $X_i$'s from $X$ yields certain error in the estimated partition. All of these reduce the fold of improvement to an amount less than $\lambda^{0.5}$ in Eq. (21). Hence, $\lambda^{0.5}$ is the maximum fold of improvement in RMSMD by exploitation of temporal correlation. For several types of available emitters[75], the average number of activations before bleaching is in the range of $\lambda \cong 30 \sim 80$ and therefore exploitation of temporal correlation can improve RMSMD by a maximum fold of $\lambda^{0.5} \cong 5.5 \sim 9$.

In an experiment, if we know the estimated locations that are produced by the same emitter, simply averaging the estimated locations per emitter can improve RMSMD by a fold as large as $\lambda^{0.5}$. However, practically only the entire set $X$ is known and its partition sets $X_i$'s are unknown and need to be estimated. To develop an algorithm that can effectively identify the partition sets $X_i$'s from the set of all estimated locations $X$ is the key to improve the quality of an FFL image through exploitation of temporal correlation.

## Results and discussion

**A numerical example.** In this section we present a numerical example to demonstrate the properties of RMSMD for the FFL images. For simplicity, we consider that infinitely many emitters are located at the grids on the entire 2D plane $\mathbb{R}^2$, $s_{ij} = (ia, ja)$ with $a > 0$ for all integers $i, j$. The Voronoi cell of $s_{ij}$ is

$$V(s_{ij}) = [ia - 0.5a, ia + 0.5a] \times [ja - 0.5a, ja + 0.5a]$$

Figure 1a shows the emitter placement for $a = 200$ nm. The estimated locations $x_{ijk} \in X_{ij} = \{x_{ij1}, \ldots, x_{ijN_{ij}}\}$ for emitter $s_{ij}$ are Gaussian distributed with mean $E(x_{ijk}) = s_{ij} + b_{ij}$ and covariance matrix $C = \mathrm{diag}(\sigma^2, \sigma^2)$.

*Invariance to a large number of estimates.* Consider that $x_{ijk}$'s are unbiased with $b_{ij} = 0$ and $\sigma \ll a$. Then all the estimated locations $x_{ijk}$'s for emitter $s_{ij}$ are almost located inside its own Voronoi cell $X_{ij} \subseteq V(s_{ij})$. By Eq. (4)

$$h(X,S) = \sqrt{2}\sigma. \tag{22}$$

As $\sigma$ increases, the RMSE between $X$ and $S$ increases without bound. However, the random RMSMD behaves quite differently. By means of Property 2

$$\lim_{\lambda\to\infty} D(X,S) = \sqrt{2}\sigma. \tag{23}$$

Figure 1b, d shows nanoscopy images $X$ for $\sigma = 25$ nm with $\lambda = 10$ and $\lambda = 25$, respectively. All estimated locations are inside the Voronoi cells of their own emitter locations and Property 2, Property 4, and Property 5 are applicable. Figure 1f shows the RMSMDs and RMSEs of $X$ and $\hat{X}$ versus $\lambda$ for $\sigma = 25$ nm. When $\lambda$ is small, $D(X,S)$ randomly varies significantly and presents a high uncertainty in quality, implying that the chance in an experiment to get a low-quality FFL image with a large RMSMD is high. As $\lambda$ increases, the variation decreases. When $\lambda = 10$, the variation is small and $D(X,S)$ is close to the expected limit value $2^{0.5}\sigma$. As predicted by Property 1, continuing to increase $\lambda$ by increasing the total number of frames $L$ and/or the activation probability $p$,

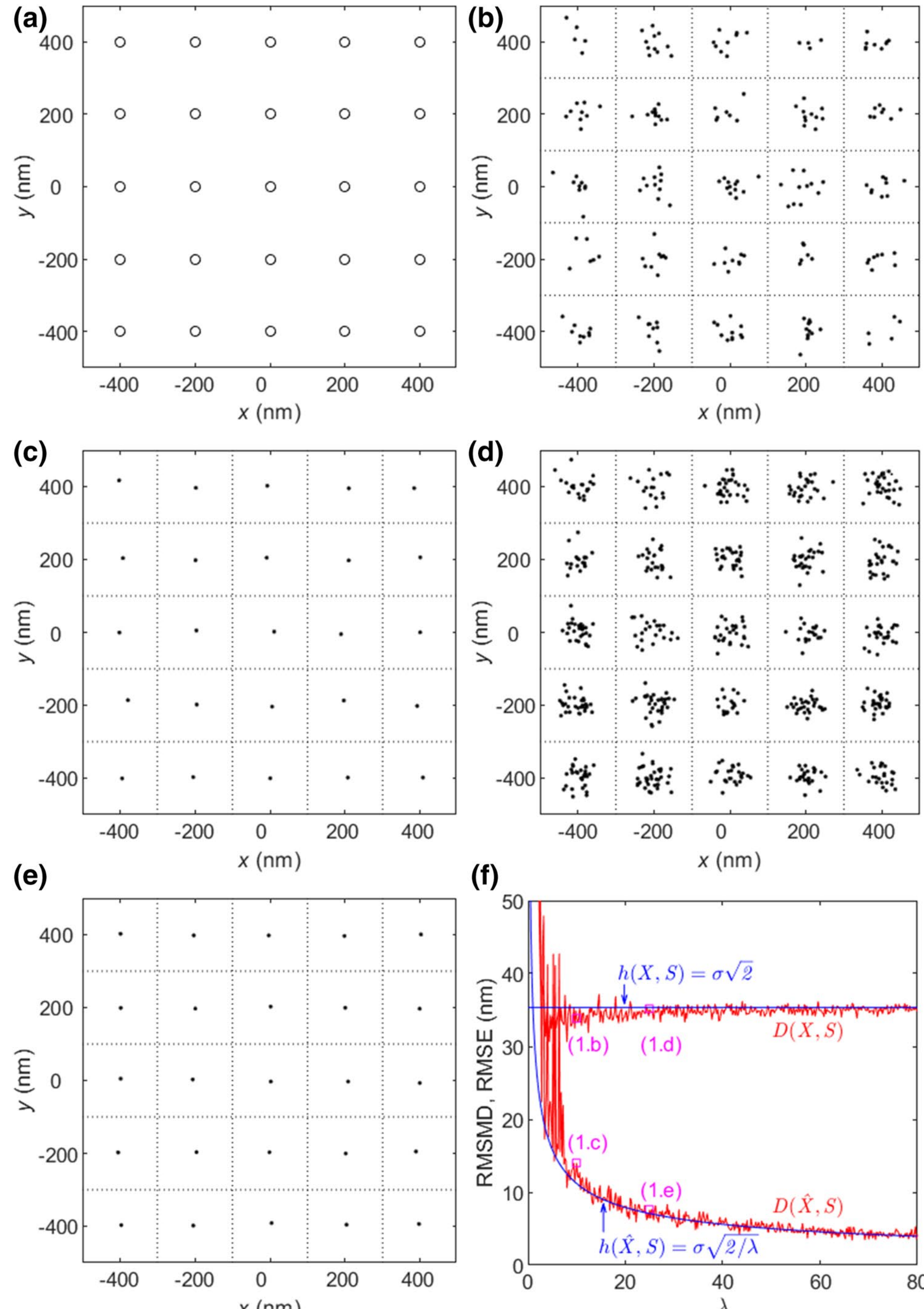

**Figure 1.** Effect of $\lambda$ and the maximum folds of RMSMD and RMSE improvements by exploitation of temporal correlation with zero bias and $\sigma = 25$ nm. (**a**) $S$ with $a = 200$ nm. (**b**) $X$ with $\lambda = 10$. (**c**) $\hat{X}$ obtained by averaging from (**b**). (**d**) $X$ with $\lambda = 25$. (**e**) $\hat{X}$ obtained by averaging from (**d**). The Voronoi cells of $s_i$'s in (**b**)–(**e**) are denoted by the dotted lines. (**f**) RMSMDs and RMSEs of $X$ and $\hat{X}$ versus $\lambda$. The RMSMDs of (1.b)-(1.e) are denoted by the squares.

only slightly reduces the variation. $D(X, S)$ eventually converges to its limit $2^{0.5}\sigma$ as predicted by Property 2. In other words, when $\lambda$ is large, the quality of all FFL images $X$'s in practice is almost the same in terms of RMSMD. As shown in Fig. 1f, the RMSMDs of Fig. 1b, d with $\lambda = 10$ and 25, respectively, differ slightly. In practice, when the average number of activations per emitter reaches $\lambda = 10$, it is unnecessary to acquire more data frames in order to reduce RMSMD variation or uncertainty of an FFL image.

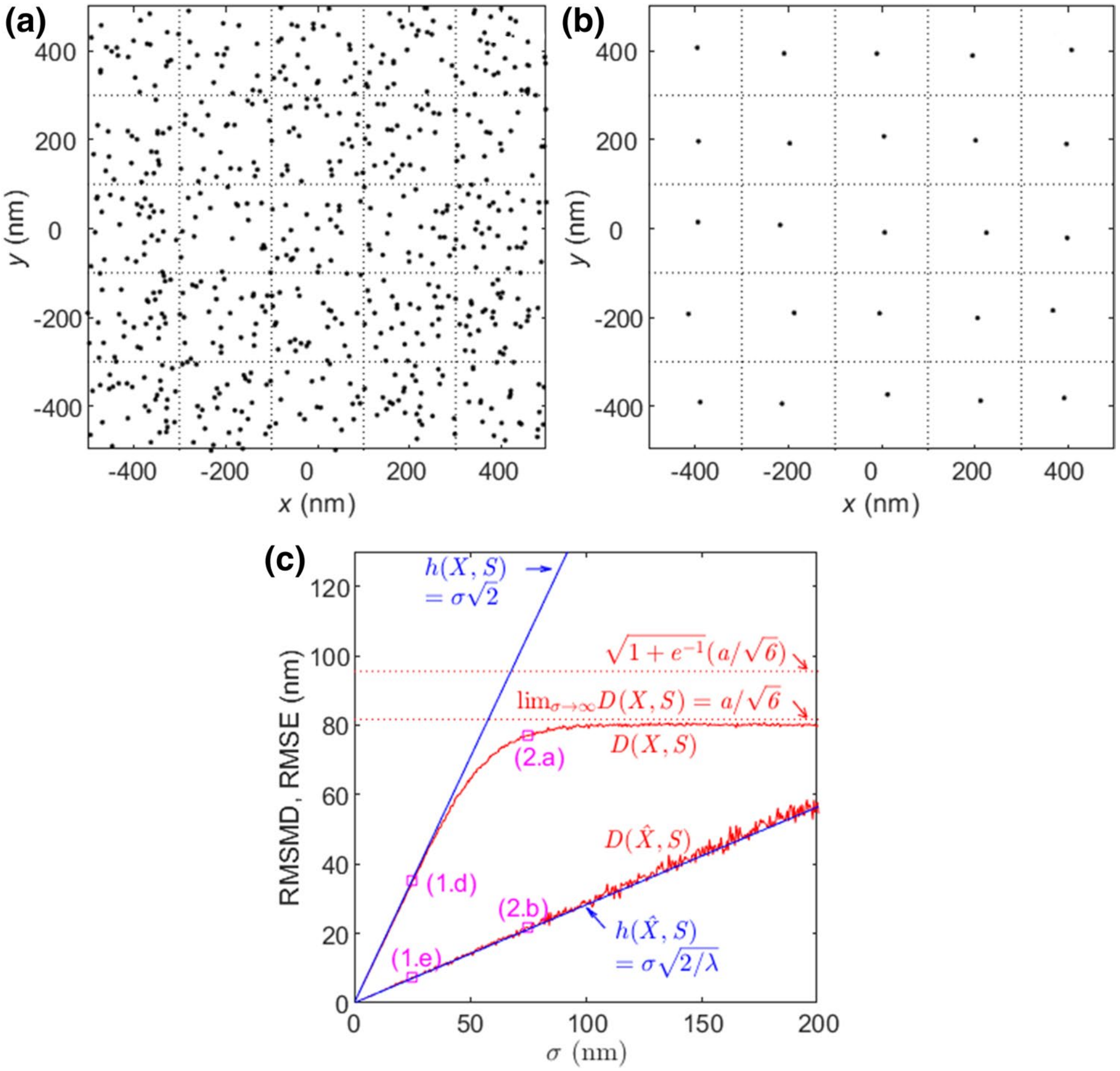

**Figure 2.** Effect on RMSMD and RMSE by large localization errors with zero bias and $\lambda = 25$. (**a**) $X$ with $\sigma = 75$ nm. (**b**) $\hat{X}$ obtained by averaging from (**a**). (**c**) The RMSMDs and RMSEs of $X$ and $\hat{X}$ versus $\sigma$. The RMSMDs of (1.d), (1.e), (2.a), (2.b) are denoted by squares.

*Averaging to reduce RMSMD.* Taking the average of the estimated locations per emitter, $\hat{x}_{ij} = N_{ij}^{-1}\sum_{k=1}^{N_{ij}} x_{ijk}$, produces a set $\hat{X}$ of estimated locations each for one emitter. For the Gaussian distribution, $\hat{x}_{ij}$ is the maximum likelihood estimator. By the symmetry of emitter placement, it follows from Eq. (11) that for the unbiased estimates

$$h(\hat{X}, S) = \sigma\sqrt{2/\lambda} \tag{24}$$

and then

$$D(\hat{X}, S) \cong \sigma\sqrt{2/\lambda} \tag{25}$$

for a large $\lambda$ and $\sigma \ll a$. Exploitation of temporal correlation reduces RMSMD by the maximum fold of $\lambda^{0.5}$. The image $\hat{X}$ in Fig. 1c is obtained by averaging the locations estimated for the same emitter in the image $X$ in Fig. 1b. After averaging, the RMSMD is reduced approximately by the maximum fold of $\lambda^{0.5} = 10^{0.5} \cong 3.2$ as shown in Fig. 1f. Moreover, with one estimated location for one emitter, the image $\hat{X}$ in Fig. 1c has a much better visual quality than that in Fig. 1b. Similarly, as shown in Figs. 1f and 2c, the RMSMD of Fig. 1e is about the maximum $\lambda^{0.5} = 5$ times lower than that of Fig. 1d, and the former presents a much better visual quality than the latter. Figure 1f shows that as $\lambda$ increases, the RMSMD and RMSE of $\hat{X}$ monotonically decrease at the rate about $\lambda^{-0.5}$ as Property 2 and Property 5 predict.

*RMSMD upper bound in a large localization error.* We investigate how RMSMD is affected by a large localization error with zero bias $b_{ij} = 0$. As $\sigma$ increases, the estimated emitter locations $x_{ijk}$ spread and RMSE in Eq. (22) monotonically increases without bound. This implies that the localization of emitters becomes worse and worse. However, as an estimated location $x_{ijk}$ enters another emitter's Voronoi cell $V(s_{lm})$, RMSMD considers only the distance between $x_{ijk}$ and the nearest emitter location $s_{lm}$ instead of its own emitter location $s_{ij}$. Consequently, RMSMD is eventually upper bounded and converges to a finite limit as $\sigma \to \infty$. In the limit, all estimated locations are uniformly distributed over the entire 2D plane. Due to the symmetric placement of $s_{ij}$ and the uniform distribution of $x_{ijk}$, the RMSMD between $X$ and $S$ over the 2D plane is equal to the RMSMD over one Voronoi cell, say the Voronoi cell of $s_{00} = (0, 0)$. According to Property 3, the limit MSMD is equal to the expectation of $\|x\|^2$ with $x$ uniformly distributed over $V(s_{00})$, that is

$$\lim_{\sigma\to\infty}\lim_{\lambda\to\infty} D^2(X, S) = a^2/6. \tag{26}$$

This implies that as RMSE increases, RMSMD is eventually upper bounded and converges to the constant $a/6^{0.5}$.

All FFL images $X$'s for a sufficiently large $\sigma$ are statistically identical and their visual qualities are also the same. The uniform distribution of estimated locations in the limit of $\sigma \to \infty$ is equivalent to a random guess of the emitter locations and no information about the emitter locations is embedded in the uniform distribution. Because of this, the limit RMSMD of $a/6^{0.5}$ is considered the upper bound of RMSMD, which correspond to the worst quality of a localization nanoscopy image.

Now we determine the limit $D(S, \hat{X})$ as $\sigma \to \infty$. For the nanoscopy image $\hat{X}$, there is one estimate $\hat{x}_{ij}$ for each emitter $s_{ij}$ and $\hat{x}_{ij}$ is uniformly distributed in the limit. Hence, the probability that $k$ estimates $\hat{x}_{ij}$'s are located in $V(s_{00})$ is a Poisson distribution with a unit mean, that is, $e^{-1}/k!$. Denote by $D_k(\hat{X}_\infty, s_{00})$ the limit RMSMD over the Voronoi cell $V(s_{00})$ that contains $k$ estimated locations. By Eq. (6), the limit $D(\hat{X}, S)$ is given by

$$\lim_{\sigma\to\infty}\lim_{\lambda\to\infty} D^2(\hat{X}, S) = \sum_{k=0}^{\infty} \frac{e^{-1}}{k!} D_k^2(\hat{X}_\infty, s_{00}). \tag{27}$$

To evaluate Eq. (27) is tedious but a lower bound can be obtained. For $k = 0$, $V(s_{00})$ contains no estimated location and the nearest $\hat{x}_{ij}$ must be outside $V(s_{00})$, and then $D_0^2(\hat{X}_\infty, s_{00}) > a^2/3$, which is the average squared distance from the origin to the boundary of $V(s_{00})$. For $k = 1$, $V(s_{00})$ contains one estimated location and $D_1^2(\hat{X}_\infty, s_{00}) = a^2/6$ given by Eq. (26). For any $k \geq 2$, we have $D_k^2(\hat{X}_\infty, s_{00}) > D_1^2(\hat{X}_\infty, s_{00})$. Therefore,

$$\lim_{\sigma\to\infty}\lim_{\lambda\to\infty} D^2(\hat{X}, S) > \left(1 + e^{-1}\right)a^2/6, \tag{28}$$

which is greater than the limit $D^2(X, S)$ in Eq. (26). This means that as the variance of localization error $\sigma^2$ increases, $D(\hat{X}, S)$ eventually surpasses $D(X, S)$ and the averaging no longer reduces RMSMD. However, as shown numerically below, this does not occur in a practical experiment where a localization error is usually much smaller than the localization error at which $D(\hat{X}, S)$ intersects with $D(X, S)$.

As shown in Fig. 2c with $\lambda = 25$, in the region of small $\sigma$, $D(X, S)$ and $D(\hat{X}, S)$ are approximately equal to $h(X, S)$ and $h(\hat{X}, S)$, respectively, and $D(\hat{X}, S)$ is improved by the maximum fold of $\lambda^{0.5} = 5$ over $D(X, S)$. As $\sigma$ increases, $D(X, S)$ is eventually upper bounded and converges to the upper bound of $a/6^{0.5}$. Meanwhile, $D(\hat{X}, S)$ increases and the improvement by the averaging dwindles. Though not shown in the figure, $D(\hat{X}, S)$ eventually surpasses $D(X, S)$ and converges to its upper bound that is slightly larger than the approximated upper bound of $(1 + e^{-1})^{0.5} a/6^{0.5}$. This means that Eq. (28) is a good approximation of the upper bound in Eq. (27), which is confirmed in Figs. 3c,d and 4d as well. Predicted by their much smaller RMSMDs, the visual qualities of images $\hat{X}$ in Figs. 1c, e and 2b are much better than those of $X$ in Figs. 1b, d and 2a, respectively, implying that the temporal correlation can significantly improve the image quality. In return, this implies that RMSMD is a rational quality metric for localization nanoscopy images. Finally, as $\sigma$ increases, $h(X, S)$ and $h(\hat{X}, S)$ increase linearly without bound, meaning that RMSE is not a proper quality metric for large localization errors.

*Effect of bias.* Now we investigate the effect of localization biases on RMSMD. Specifically, the estimated locations $x_{ijk}$ are Gaussian distributed with mean $E(x_{ijk}) = s_{ij} + b_{ij}$, $b_{ij} \neq 0$, and covariance matrix $C = \mathrm{diag}(\sigma^2, \sigma^2)$. Unlike the sample drift, the biases of estimated locations $x_{ijk}$'s for different emitter $s_{ij}$'s are usually different. To simplify the analysis, we consider that the biases $b_{ij}$ for different $i, j$ are realizations of a Gaussian random vector with mean zero and covariance matrix $\mathrm{diag}(\delta^2, \delta^2)$. This implies that as $M \to \infty$, $M^{-1}\sum_{i=1}^{M} b_{ik}^2 \to \delta^2$ almost surely. By Eqs. (4) and (11), $h^2(X, S) = 2(\sigma^2 + \delta^2)$, and $h^2(\hat{X}, S) = 2(\sigma^2/\lambda + \delta^2)$, respectively. The MSEs increase without bound as the variance of bias $\delta^2$ increases.

When both the variance of localization error $\sigma^2$ and the variance of biases $\delta^2$ across emitters are small such that the estimated locations all are almost surely located in the Voronoi cells of their own emitters, Property 2 and Property 4 are applicable and

$$D^2(X, S) = 2(\sigma^2 + \delta^2), \tag{29}$$

$$D^2(\hat{X}, S) = 2(\sigma^2/\lambda + \delta^2). \tag{30}$$

As expected, the averaging cannot reduce the effect of biases.

As $\sigma \to \infty$, $D^2(X, S)$ still converges to the right-hand side of Eq. (26) regardless of bias $\delta$. On the other hand, given $\sigma$, as $\delta \to \infty$, all estimated locations are eventually uniformly distributed, that is, $\delta$ plays a similar role in $D(X, S)$ and $D(\hat{X}, S)$ as $\sigma$ does. Hence, similar to Eqs. (26) and (28), we obtain

$$\lim_{\delta\to\infty}\lim_{\lambda\to\infty} D^2(X, S) = a^2/6, \tag{31}$$

$$\lim_{\delta\to\infty}\lim_{\lambda\to\infty} D^2(\hat{X}, S) > (1 + e^{-1})a^2/6. \tag{32}$$

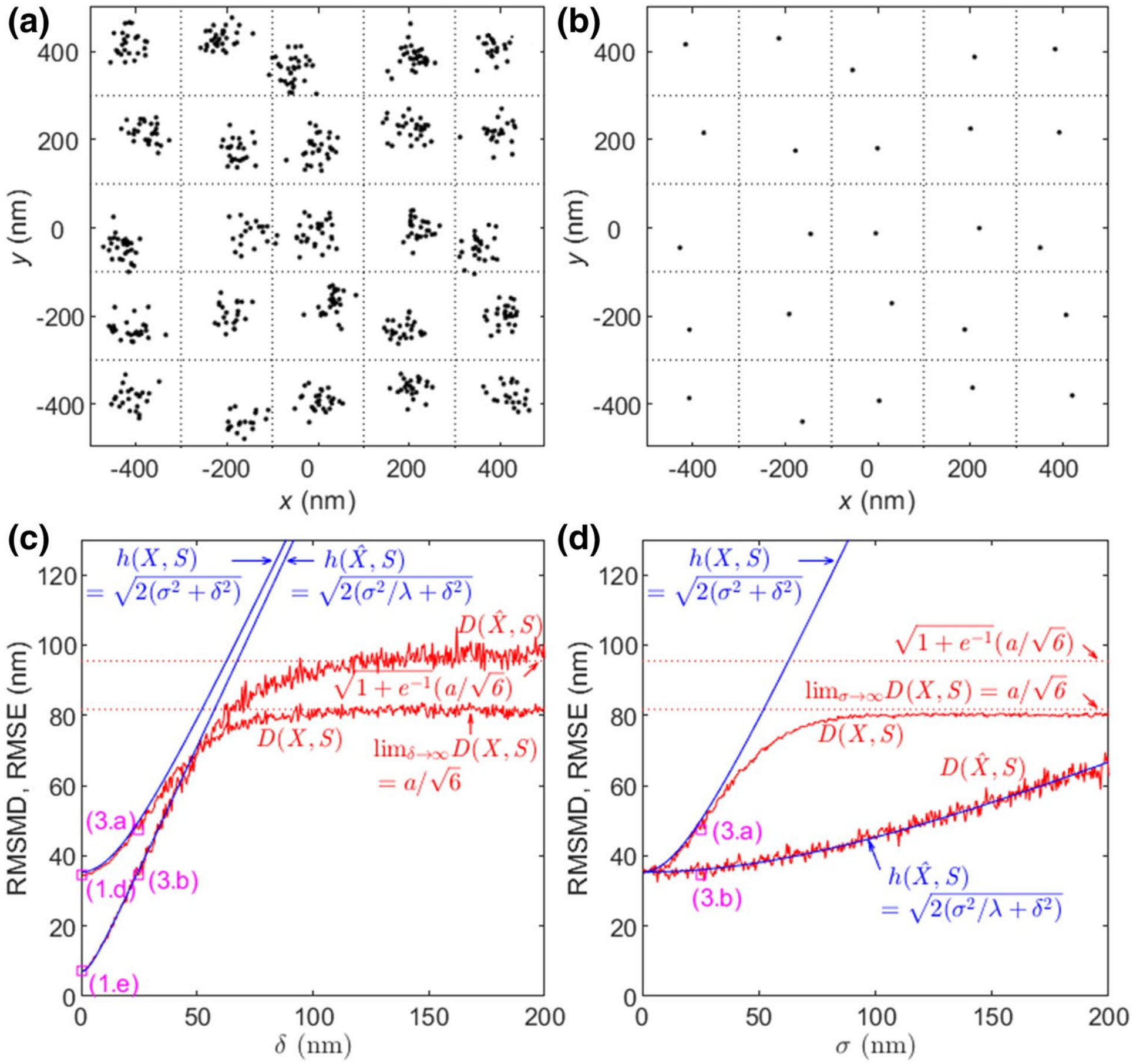

**Figure 3.** Effect of bias and variance of localization errors on RMSMD and RMSE with $\lambda = 25$. (**a**) $X$ with $\sigma = 25$ nm and $\delta = 25$ nm. (**b**) $\hat{X}$ obtained by averaging from (**a**). (**c**) RMSMDs and RMSEs of $X$ and $\hat{X}$ versus $\delta$ with $\sigma = 25$ nm. The RMSMDs of (1.d), (1.e), (3.a), and (3.b) are denoted by squares. (**d**) RMSMDs and RMSEs of $X$ and $\hat{X}$ versus $\sigma$ with $\delta = 25$ nm. The RMSMDs of (3.a) and (3.b) are denoted by squares.

Figure 3 shows the effect of bias and variance of localization errors on RMSMD and RMSE with $\lambda = 25$. Figure 3a,b respectively show an image of $X$ and its corresponding $\hat{X}$ with $\sigma = 25$ nm and $\delta = 25$ nm. The biases for different emitters are different. As shown in Fig. 3c, in the region of small $\delta$, the RMSMDs of $X$ and $\hat{X}$ are approximately equal to $h(X, S)$ and $h(\hat{X}, S)$, respectively. As $\delta$ increases, both eventually diverge and vary significantly around their upper bounds in Eqs. (31) and (32), respectively. As $\sigma$ increases, the RMSMDs of $X$ and $\hat{X}$ with $\delta = 25$ nm behave similarly to those in Fig. 2c with zeros bias; but the former is lifted and pressed towards the upper bounds due to the bias. It is worthy to point out that comparing Fig. 3c, d, the RMSMDs of both $X$ and $\hat{X}$ converge faster to their bounds as $\delta$ increases than as $\sigma$ increases. Moreover, the biases cannot be reduced by the averaging. Therefore, the variance of biases across emitters affects more severely on the RMSMD than the variance of localization errors.

*Effect of sample drift.* We investigate the effect of sample drift on RMSMD. While the localization errors cause different biases across emitters, a sample drift produces the same bias on all emitters. Because of this, the drift in a localization nanoscopy image is easy to identify and remove. Nevertheless, the effect of a drift on RMSMD is significant as analyzed below.

Consider a sample drift $(d_1, d_2)$ and that estimated locations $x_{ijk}$'s are Gaussian distributed with mean $E(x_{ijk}) = s_{ij} + (d_1, d_2)$ and covariance matrix $C = \mathrm{diag}(\sigma^2, \sigma^2)$. By Eqs. (4) and (11), the MSEs of $X$ and $\hat{X}$ are equal to $h^2(X, S) = 2\sigma^2 + d_1^2 + d_2^2$ and $h^2(\hat{X}, S) = 2\sigma^2/\lambda + d_1^2 + d_2^2$, respectively. The MSEs increase without bound as the drift increases.

In comparison, as the drift increases, the $X$ with the drift of $(d_1, d_2)$ is statistically identical to the $X$ with the drift of $(d_1, d_2) + (ma, la)$ for integers $m$ and $l$. This implies that $D(X, S)$ varies periodically with a period of $a$ as the drift changes, and so does $D(\hat{X}, S)$. Considering the period, when the estimated locations for one emitter are all located inside the Voronoi cell of an emitter, the RMSMDs of $X$ and $\hat{X}$ are still determined by Property 2 and Property 4, that is,

$$D^2(X, S) = 2\sigma^2 + d_1^2 + d_2^2, \tag{33}$$

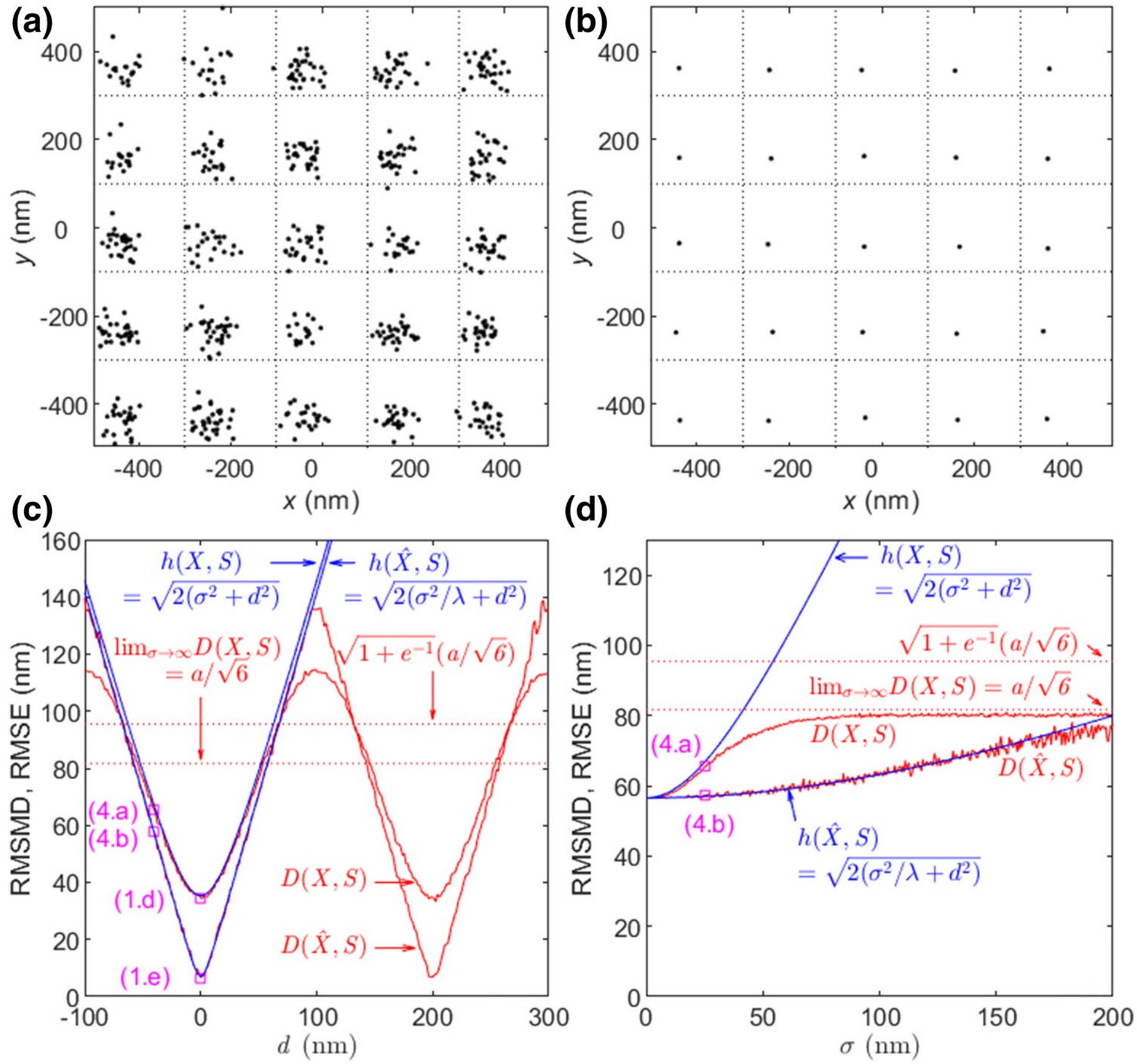

**Figure 4.** Effect of sample drift on RMSMD and RMSE with $\lambda = 25$. (**a**) $X$ with $\sigma = 25$ nm and $d = -40$ nm. (**b**) $\hat{X}$ obtained by averaging from (**a**). (**c**) RMSMDs and RMSEs of $X$ and $\hat{X}$ versus $d$ with $\sigma = 25$ nm. The RMSMDs of (1.d), (1.e), (4.a), and (4.b) are denoted by squares. (**d**) RMSMDs and RMSEs of $X$ and $\hat{X}$ versus $\sigma$ with $d = -40$ nm. The RMSMDs of (**a**), and (**b**) are denoted by squares.

$$D^2(\hat{X}, S) = 2\sigma^2/\lambda + d_1^2 + d_2^2. \tag{34}$$

The averaging does not change the sample drift. As $\sigma \to \infty$, the RMSMDs of $X$ and $\hat{X}$ still converge respectively to their upper bounds of Eqs. (26) and (27) regardless of drift.

In Fig. 4a–d the drift $(d_1, d_2) = (d, d)$ is considered with $\lambda = 25$. Figure 4a is an image of $X$ with $\sigma = 25$ nm and $d = -40$ nm in which the effect of drift can be seen. Figure 4b is the image $\hat{X}$ obtained by averaging from (a). The visual quality is significantly improved and the effect of sample drift can be seen unchanged. The RMSMDs of (a) and (b) are indicated in (c) and (d).

Figure 4c demonstrates how RMSMD and RMSE change with respect to $d$ with $\sigma = 25$. As $|d|$ increases, RMSMDs of $X$ and $\hat{X}$ are eventually bounded and vary periodically with the period of $a$ while RMSEs increase without bound. When the drift is small such as $d = -40$ nm in (**a**) and (**b**), $D(X, S) \cong h(X, S)$ and $D(\hat{X}, S) \cong h(\hat{X}, S)$. However, when the drift is large, they diverge significantly. It is noticeable that the worst drift is $d = (k + 0.5)a$ such that the estimated locations are located at the vertices of four adjacent Voronoi cells. In this case, RMSMDs of $X$ and $\hat{X}$ are larger than the corresponding bounds in Eqs. (31) and (32). However, under any sample drift the nanoscopy image has the same pattern as that after the drift is removed. Furthermore, a sample drift is easy to identify and remove. Hence, the upper bounds in Eqs. (26) and (27) are still considered the highest RMSMD that correspond to the worst quality of nanoscopy images where no information of emitter locations is contained.

Figure 4d demonstrates how RMSMD and RMSE change with respect to $\sigma$ with the drift of $d = -40$ nm. The RMSMDs of $X$ and $\hat{X}$ behave similarly to Fig. 3d with bias. Due to the symmetric placement of emitters and the small drift, the effect of drift is similar to that of a bias.

**Methods to exploit temporal correlation.** In this paper, we focus on analysis of the potential maximum quality improvement of FFL images by exploiting its embedded temporal correlation. In this section, we briefly discuss how to develop such an algorithm to exploit the temporal resolution. First, due to its photodynamics each emitter is usually assumed to be activated independently from one frame to another in accordance with a Markov chain[44, 45, 66]. The estimated locations of the same emitter in an FFL image are presented following the Markov chain and therefore are temporally correlated. The properties of the Markov chain can be exploited to identify the estimated locations that are generated by the same emitter. Second, the estimated locations of an emitter are usually distributed as a cluster around its true location. Exploiting the property of clusters can detect the identity of estimated locations and improve the image quality. Many existing clustering algorithms in the literature[76] are ready to apply, say the k-means method, the methods based on the Gaussian mixture model, etc. While an algorithm can exploit either the property of a Markov chain or the property of clusters, an optimum algorithm shall jointly exploit the properties of both the Markov chain and the clusters, which is necessarily more computationally complex.

## Conclusions

We have analyzed the statistical properties of root mean square minimum distance (RMSMD) and root mean square error (RMSE) for the frame-by-frame localized (FFL) nanoscopy images. When the average number of estimated locations per emitter $\lambda$ is greater than ten, the variation of RMSMD is slightly reduced by continuously increasing $\lambda$; and then increasing the number of data frames in an acquisition is unnecessary. On the other hand, exploiting the temporal resolution can reduce RMSMD and RMSE by a maximum fold of $\lambda^{0.5}$ and therefore the fold of improvement increases as the number of acquired data frames increases. When the localization error is small, the RMSMD and the RMSE coincide. As the localization error increases without bound, the RMSE increases without bound; in contrast, the RMSMD is eventually upper bounded by that of the worst nanoscopy image where all estimated locations are uniformly distributed and no information about emitter locations is contained. This implies that even for large localization errors, the RMSMD is a proper metric of image quality but the RMSE is not without saying that the RMSE is not applicable in practice. The random biases of localization errors across emitters affect the RMSMD in the similar way to the variance of localization errors but the former affects more severely than the latter. As the sample drift increases, the RMSMD goes up and down alternately. The analytical results for the emitters located on the 2D grids can be used as a reference to benchmark the quality of FFL nanoscopy images. An algorithm to exploit the temporal resolution can take use of the properties of the Markov chain and clusters. The results suggest to develop two kinds of algorithms: the algorithms that can exploit the temporal correlation of FFL nanoscopy images and the unbiased localization algorithms.

## Appendix

***Proof of Eq. (4)*** Since all $x_{ij} \in X_i$ follow the same probability density function $f_i(x)$, the MSE of Eq. (3) can be written as

$$h^2(X,S) = \frac{1}{M}\sum_{i=1}^{M} E(\|x_{i1} - s_i\|^2)$$

$$= \frac{1}{M}\sum_{i=1}^{M}\sum_{k=1}^{n} E[(x_{i1k} - s_{ik})^2]. \tag{35}$$

Since $E[(x_{i1k} - s_{ik})^2] = E[(w_{i1k} + b_{ik})^2] = \sigma_{ik}^2 + b_{ik}^2$, the MSE of $X$ is given by Eq. (4). □

***Proof of Property 1*** As $\lambda \to \infty$, both $N_i \to \infty$ and $N \to \infty$ in probability[77]. In Eq. (1) each term $\min_{x\in X}\|x - s\|^2$ in the first sum must be included in the second sum. Moreover, there are about $\lambda$ times more terms in the second sum than in the first sum. The first sum is infinitesimal in the limit. It follows from Eq. (1) and the law of large numbers that

$$\lim_{\lambda\to\infty} D^2(X,S) = \lim_{\lambda\to\infty}\frac{1}{N}\sum_{x\in X}\min_{s\in S}\|s - x\|^2 \tag{36}$$

$$= \lim_{\lambda\to\infty}\frac{1}{N}\sum_{i=1}^{M}\sum_{j=1}^{N_i}\min_{s\in S}\|s - x_{ij}\|^2$$

$$= \lim_{\lambda\to\infty}\sum_{i=1}^{M}\frac{N_i}{N}\frac{1}{N_i}\sum_{j=1}^{N_i}\min_{s\in S}\|s - x_{ij}\|^2$$

$$= \lim_{\lambda\to\infty}\sum_{i=1}^{M}\frac{N_i}{N}E\left(\min_{s\in S}\|s - x\|^2|X_i\right) \tag{37}$$

$$= E\left[E\left(\min_{s\in S}\|s - x\|^2|X_i\right)\right]$$

which yields Eq. (5). Equation (37) holds almost surely since there are $N_i$ locations $x_{ij} \in X_i$ and the expectation in Eq. (37) is taken with the condition of $x \in X_i$.

Since $x_{ij}$'s have a stationary probability density function $f_i(x)$, Eq. (5) can be further written as

$$E\left(\min_{s\in S}\|s-x\|^2\right)=\frac{1}{M}\sum_{i=1}^{M}\frac{1}{N_i}\sum_{j=1}^{N_i}E\left(\min_{s\in S}\|s-x_{ij}\|^2\right)$$

$$=\frac{1}{M}\sum_{i=1}^{M}E\left(\min_{s\in S}\|s-x_{i1}\|^2\right). \tag{38}$$

By means of the Voronoi cells of $s_i$'s, Eq. (38) can be expressed as

$$E\left(\min_{s\in S}\|s-x\|^2\right)=\frac{1}{M}\sum_{i=1}^{M}\sum_{j=1}^{M}\int_{V(s_j)}\|s_j-x\|^2 f_i(x)dx \tag{39}$$

which is equal to Eq. (6). □

***Proof of Property 2*** With the given condition, we have

$$\int_{V(s_i)} f_j(x)dx=\delta_{ij}$$

for all $i$, $j$, and then

$$\int_{V(s_i)}\|s_i-x\|^2 f_i(x)dx=\int_{R^n}\|s_i-x\|^2 f_i(x)dx,$$

$$\int_{V(s_i)}\|s_i-x\|^2 f_j(x)dx=0.$$

It follows from Eq. (6) that

$$\lim_{\lambda\to\infty}D^2(X,S)=\frac{1}{M}\sum_{i=1}^{M}\int_{R^n}\|s_i-x\|^2 f_i(x)dx \tag{40}$$

$$=\frac{1}{M}\sum_{i=1}^{M}E(\|s_i-x_{i1}\|^2)$$

$$=\frac{1}{M}\sum_{i=1}^{M}\sum_{k=1}^{n}(\sigma_{ik}^2+b_{ik}^2). \tag{41}$$

Hence, Eq. (8) holds due to Eq. (4). □

***Proof of Eq. (11)*** It is clear that the mean of $\hat{x}_i$ is $E(\hat{x}_i)=s_i+b_i$ and the variance of its $k$ th component with a fixed $N_i$ is equal to $\hat{\sigma}_{ik}^2=\sigma_{ik}^2/N_i$. It follows from Eq. (35) that

$$h^2(\hat{X},S)=\frac{1}{M}\sum_{i=1}^{M}\sum_{k=1}^{n}E[(\hat{x}_{ik}-s_{ik})^2]$$

$$=\frac{1}{M}\sum_{i=1}^{M}\sum_{k=1}^{n}\left[E\left(\frac{\sigma_{ik}^2}{N_i}\right)+b_{ik}^2\right] \tag{42}$$

which yields Eq. (11) since $\lambda/N_i\to 1$ almost surely as $\lambda\to\infty$. □

***Proof of Property 4*** By the law of large numbers, as $\lambda\to\infty$, $\hat{x}_i\to s_i$ almost surely and then the probability that $\hat{x}_i$ is in the Voronoi cells of other $s_j$ for $j\neq i$ tends to zero, that is, $\Pr(\hat{x}_i\in V(s_j))\to\delta_{ij}$ for all $i$, $j$; and correspondingly, $\Pr(s_i\in V(\hat{x}_j))\to\delta_{ij}$. $\hat{X}$ and $S$ are a pair of kernel sets. It follows from Eq. (3) in Ref.[66] that in the almost sure sense

$$\lim_{\lambda\to\infty} D^2(\hat{X}, S) = \lim_{\lambda\to\infty} \frac{1}{M}\sum_{i=1}^{M} \|\hat{x}_i - s_i\|^2.$$

Since $\hat{\sigma}_{ik}^2 = \sigma_{ik}^2/N_i$, $\hat{x}_i \to s_i + b_i$. Hence, almost surely

$$\lim_{\lambda\to\infty} D^2(\hat{X}, S) = \frac{1}{M}\sum_{i=1}^{M} \|b_i\|^2, \tag{43}$$

which yields Eq. (13) in terms of Eq. (12). □

## Acknowledgements

This work was partly supported by a PSC-CUNY Award, jointly funded by The Professional Staff Congress and The City University of New York.

## Author contributions
Y.S. contributed the entire work of this manuscript.

## Competing interests
The author declares no competing interests.

## Additional information
**Correspondence** and requests for materials should be addressed to Y.S.

**Reprints and permissions information** is available at www.nature.com/reprints.

**Publisher's note** Springer Nature remains neutral with regard to jurisdictional claims in published maps and institutional affiliations.